# Investigating Modes of Activity and Guidance for Mediating Museum Exhibits in Mixed Reality

Katrin Glinka
Humboldt-Universität zu Berlin · 10117 Berlin
katrin.glinka@hu-berlin.de

Patrick Tobias Fischer, Claudia Müller-Birn
Freie Universität Berlin · 14195 Berlin
patrick.tobias.fischer@fu-berlin.de, clmb@inf.fu-berlin.de

Silke Krohn
Stiftung Preußischer Kulturbesitz · 10709 Berlin
s.krohn@smb.spk-berlin.de

**Abstract**

We present an exploratory case study describing the design and realisation of a "pure mixed reality" application in a museum setting, where we investigate the potential of using Microsoft's HoloLens for object-centred museum mediation. Our prototype supports non-expert visitors observing a sculpture by offering interpretation that is linked to visual properties of the museum object. The design and development of our research prototype is based on a two-stage visitor observation study and a formative study we conducted prior to the design of the application. We present a summary of our findings from these studies and explain how they have influenced our user-centred content creation and the interaction design of our prototype. We are specifically interested in investigating to what extent different constructs of initiative influence the learning and user experience. Thus, we detail three modes of activity that we realised in our prototype. Our case study is informed by research in the area of human-computer interaction, the humanities and museum practice. Accordingly, we discuss core concepts, such as gaze-based interaction, object-centred learning, presence, and modes of activity and guidance with a transdisciplinary perspective.

# 1    Introduction[1]

As in other areas of society, there is a growing interest within cultural heritage institutions to expand existing formats of interpretation and communication by means of digital technologies. With the case study presented in this article, we contribute to this growing interest in digital mediation efforts by exploring how we can conceptualise and design a mixed reality application with a transdisciplinary approach that draws from museum discourses and human-computer interaction. We are particularly interested in two aspects of museum mediation: object-centredness and guidance. Our case study is based on the concept of the "trained gaze", which we seek to translate into a "digitally augmented gaze" for non-expert museum visitors. More specifically, we explore how a visitor can independently examine a museum object in a mixed reality experience while the system supports the visitor in deciphering the visual characteristics of the object and, thus, reveals which properties form the basis for understanding the symbolic or historical meaning of the object. The case study is motivated by three leading research interests, which we investigate with our prototype in an *in situ* user study at a later stage.[2] Firstly, we are interested to see whether and to what extent visitors perceive the "presence" of the physical object exhibited, even if it is supplemented with digital content and virtual objects. Secondly, we adapt different mediation strategies to implement a "digitally augmented gaze" in three modes of activity: a guided narrative, a self-guided exploration mode and a co-active, bidirectional format, in which the system and the visitor can take the initiative. We are particularly interested in the extent to which these modes of activity and guidance affect the learning success of visitors. Finally, we aim to determine whether and how the three different modes affect the overall user experience.

1    Acknowledgements: This case study was conducted as a collaboration between the Freie Universität Berlin (Patrick Tobias Fischer and Claudia Müller-Birn) and the Prussian Cultural Heritage Foundation (Katrin Glinka and Silke Krohn) as part of museum4punkt0, a project funded by the Federal Government Commissioner for Culture and Media. We wish to thank the Deutsches Historisches Museum for contributing the study object, technical infrastructure and support, and additional material and content. We also thank Anna Heib and Marie-Christin Fahs for their valuable assistance in conducting the visitor study. The programming of the prototype was finalised by FRAMEFIELD.

2    Due to the closure of museums caused by COVID-19, it was not possible for us to conduct the in situ user study that we had planned for March 2020 before finishing this paper.

In the following, we first lay out our interpretation strategy of augmenting a "trained gaze" and contextualise it in related museum discourses. We then describe the considerations we derived from this concept for the design and technical realisation. This is followed by a discussion, firstly, of object-centredness and object presence in a museum context and in human-computer interaction and, secondly, of different modes of activity and guidance in museum mediation. The second part of the paper presents the findings from a two-stage visitor observation study and formative study that are part of a user-centred approach that informed the content creation and development. In the third and final part of the paper, we describe systematically how research insights from human-computer interaction have informed the realisation of our concept of augmenting a physically present museum object in three different modes of activity. With this paper, we want to illustrate a transdisciplinary approach for the design and development of interactive digital media, especially mixed reality applications in museum contexts. We conclude our article with a discussion on the future steps needed.

## 2      Interpretation Strategy and Design Considerations

Museum mediation promotes understanding and meaning-making among visitors. Although there are different understandings of what exactly is meant by the term *mediation* – depending on the country, cultural institution and tradition of practice – the term generally covers three interconnected aspects: education, interpretation and communication [Medi14: 20]. *Interpretation*, in this context, refers to a dynamic process that aims at assisting visitors to develop the skills to explore an exhibition or object independently and enhance their understanding [Blac05: 185]. The term *interpretation* is normally used to describe informal, voluntary and enjoyable learning in museums that caters to the interest of the visitor [ibid.: 183], whereas *museum education* refers to rather formal learning situations.

Museums make use of a variety of media, in both analogue and digital formats, to offer interpretation about objects. However, most interpretive media require the visitor to avert their eyes away from the objects exhibited, at least to some extent, be it to read wall texts, handouts, catalo-gues, labels, to interact with digital displays, or even during guided tours, where people might focus on the guide or engage in other social interactions with their group. Even with audio guides, which allow visitors to listen to additional information while viewing exhibits or strolling through a museum, visitors have to shift their focus of attention away from the objects in order to select the next audio clip on their device. Furthermore, the systems used for delivering the audio content have no way of detecting whether the visitor is actually looking at the object for which it is providing additional content. Based on these considerations, we decided to develop an object-centred interpretation strategy that minimises the need to avert the eyes away from the object while using a system that allows us to directly connect interpretation to the visual properties of the object at which a visitor is currently looking.

We took inspiration for our prototype from the concept of a "trained gaze" or of acquiring "viewing skills", which include observational skills, the ability to probe and to find a variety of possible meanings [RiYe02]. By building our interpretation strategy on the idea of a "trained gaze", we take into account the assumption that a work of art can only be truly understood and grasped through precise and prolonged observation [teHe12: 15]. However, this assumption presupposes a high degree of expertise and connoisseurship and can only apply if the viewer has the necessary art historical and iconographic knowledge to "read" an object and its formal properties [Dudl12a: 11]. Art historians rely, among other things, on a "mental archive of images and style" [HaKl06: 12] when observing an artwork. This incorporated implicit and explicit knowledge combined with acquired viewing skills are part of a "trained gaze". The latter does not only influence the ability to derive meaning from the act of visually observing objects, but it also influences the perception and aesthetic judgement of art [NoLK93].

Regarding the realisation of this interpretation strategy, we focused on three main principles and considerations for our choice of mediation technology and interaction design that we present in the following sections: gaze-based interaction in a pure mixed reality experience (2.1), object-centredness and a sense of object-presence (2.2), and different modes of activity and guidance (2.3).

## 2.1 Gaze-based Interaction in Pure Mixed Reality

With our interpretation strategy in mind, our choice of technology has to allow thorough observation of the museum artefact without obstructing the view of the visitor. Moreover, we seek to minimise the need to avert the eyes away from the object while visually exploring it. Taking these requirements into account, we decided to explore the potentials of mixed reality, which led to our choice of Microsoft's HoloLens [MS20a], a head-mounted holographic see-through projection system with a mobile processing unit and various sensors. In contrast to conventional augmented reality technology, these glasses enable a true merge of digital content with the physical world, rather than superimposing it. Carlos Flavian et al. [FlIO18] call this type of mixed reality "a pure mixed reality" (PMR) and position this concept between augmented reality and augmented virtuality in Milgram and colleagues' "Virtuality Continuum" [MTUK94]. Visitors are enabled to equally and simultaneously interact with virtual and physical objects in a PMR.

In the context of the HoloLens technology, we considered three main aspects for the realisation of our PMR application. Firstly, the merging of physical and virtual objects is made possible through the *spatial mapping* feature, which provides a detailed real-time representation of the physical surfaces in the environment around the HoloLens. This so-called "world mesh" enables the designer of a PMR environment to interweave physical objects convincingly with the virtual space. *Inter-reality physics* becomes possible, as virtual objects can, for example, bounce off tables, floors, chairs, or other spatial or physical elements that have been integrated into the world model of the HoloLens. Interaction techniques based on the spatial mapping feature, respectively, on inter-reality physics, can contribute to a strong sense of PMR.

The *sliding puck*, a "cursor" that is attached to the user's gaze, is the second important aspect in the design of our PMR. In combination with the *world mesh*, the *sliding puck* behaves equally to virtual objects as it does to physical objects. The puck slides over the surface of the object of which the system has created a world mesh (see Fig. 1). While the users let their gaze wander, the puck simultaneously aligns with the surface of the objects mapped. The users, metaphorically speaking, can "touch" the objects with the sliding puck. Physical and digital objects can be treated equally in a PMR, since the gaze aligns with the surface of virtual objects in the same way as it does with physical objects. This offers the opportunity, for example, to virtually augment missing elements of an object while treating both object surfaces, the physical and the virtual, equally from an interaction point of view.

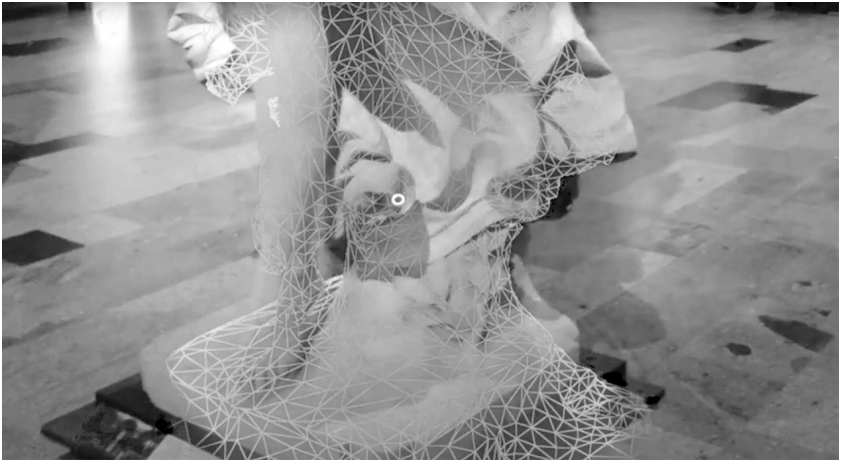

Fig. 1: HoloLens' spatial mapping feature creates a "world mesh" of its physical surroundings. The "sliding puck" (centre) aligns to the surface of all objects included in this world mesh – both existing physical objects and virtual objects that are added to the world mesh.

The third aspect which can contribute to a sense of being in a PMR is the *"unbroken" holospace*. This relates to the limited field of vision in the HoloLens. Especially large virtual artefacts that surpass the field of vision emphasise the boundaries of the holospace and might break the illusion of the PMR. To avoid the impact of a "broken" holospace, virtual objects can be designed rather small or distant. However, to mitigate this restriction, interaction techniques such as the "tag-along" technique [MS20b], which is a feature that lets an object follow the gaze axis, can be employed to create the illusion of a larger holospace. Also gestalt rules such as reification can be used to avoid exposing the edges of the holo field of vision.

## 2.2   Object-centred Learning and Sense of Object-Presence

As has already been implied with our interpretation strategy, our use case is focused on museums of art and cultural history. Regarding this type of museum, special attention must be paid to the significance of the original object – more precisely, the original artwork or the authentic historical artefact. Objects act as material witnesses that help us to understand art, cultures and the past. Consequently, the original object is at the centre of educational practice, especially in museums of art and cultural history [RiYe02].

Object-centredness relates to the concept of *presence*, which, in our case, needs to be discussed from two perspectives: its meaning in a museum context and human-computer interaction and the design of virtual or mixed reality. Object-centredness *per se* in museology and the humanities does not automatically invoke a preoccupation with the material presence of an object. Gumbrecht, for example, points out that from an epistemological standpoint, the core practice of the humanities constitutes an institutionalised tradition of interpretation, understood as "the identification and/or attribution of meaning" [Gumb04: 1]. However, he calls for a "reconfiguration of some of the conditions of knowledge production" (ibid.). He suggests that we instead "conceive of aesthetic experience as an oscillation (and sometimes an interference) between 'presence effects' and 'meaning effects'" [Gumb04: 2], which emphasises and acknowledges the influence of materialities on our experience and understanding of the world. We explore in our case study whether and how museum interpretation, while in itself focused on the attribution of meaning, can also simultaneously highlight the material presence of an object. This approach is set against the observation that objects in museums often "feature as mere illustrations punctuating the story being told" [Dudl12a: 6], which, in turn, distances objects from visitors by putting a focus on "the story *overlying* the physical thing" [Dudl12b: 10).

At the same time, the term *presence* also plays an important role in research on virtual, augmented and mixed reality. Although in this context, the term relates more to the feeling of "being present" or fully submerged in a virtual, augmented or mixed environment. Presence is understood as the outcome or purpose of immersion [ScFR01]. Research in this area, among other things, aims at understanding how a sense of presence emerges in such environments and how the sense of presence (or lack thereof) can be measured [UCAS00]. Furthermore, discussions of numerous types of presence have emerged since then. Exemplarily, mixed reality, specifically, aims at creating a "co-presence", a sense of being together in a shared space [WBJK09].

Our approach combines these two concepts of *presence*: the *material presence* of an object and its effects on understanding and meaning-making, and a sense of *co-presence* with the object in a mixed reality environment. We aim at creating a PMR that brings the visitor and the museum object together into a shared space in which the attention of the visitor is directed towards the material presence of the object. As Wagner et al. point out, the experience and, thus, perception of presence is created by a co-constructed reality that is "filtered through the purpose of the actions in which we are engaged" [WBJK09: 273]. This aspect also relates to and emphasises the importance of active visual exploration, which is implied in our interpretation strategy, in a mixed reality setting. Similarly, Schubert et al. point out that attention allocation and the awareness of possible action patterns both contribute to the emergence of presence. They argue "that a virtual environment, like any other environment, is perceived and understood by mentally combining potential patterns of action" [ScFR01: 267].

### 2.3   Modes of Activity and Guidance

The scope of action for visitors in museums is usually structured along two main conditions: interpretation is either conveyed in a guided or a self-guided mode. Guided, in this context, especially includes tours through an exhibition or museum where visitors are accompanied by a museum guide. These guided tours often follow a linear narrative that highlights certain aspects or perspectives on a given topic. Here, the museum guide takes on the active part, decides on a narrative and sets the route and timing, while the visitors follow rather passively. Apart from interpersonal guided formats, audio guides can be used during a visit, or an exhibition design might suggest the visitors follow a predefined path through a curatorial narration. However, even when curators or educators establish such predefined paths through an exhibition, there is evidence that suggests that visitors sometimes assert their own agency and, thus, explore exhibitions by way of their own initiative and interest; they "go where they want to go. They skip elements, visit only one-third of them, and spend much less time than is often assumed" [Serr97: 120].

A museum visit in a completely unguided modus, without resorting to any interpretative media, would presuppose that a visitor can draw from previously acquired viewing skills and a "trained gaze". However, since our interpretation strategy aims at digitally augmenting such a "trained gaze" for *non-expert* visitors, we assume that a minimum of guidance and instruction during the observation and viewing process would be needed. Such a condition can also be called "self-guided". During a self-guided tour, the visitors decide independently what they want to see or how much time they want to spend with what activity, while they receive interpretation through, among others, wall texts, labels or leaflets. Most audio or multimedia guides can also be used in a self-guided mode, which means that visitors would not use them to follow a predefined narrative through an exhibition, as in a guided format, but rather to access information about individual artefacts when desired. There has been a shift in museum mediation practice, especially in recent years, that has led to a diversification of formats and activities in museums. This includes low-threshold educational formats such as interactive games, or inclusive media such as multimedia apps in sign language, but also a variety of participatory formats that invite visitors to actively contribute to the narrative of an exhibition. This exemplifies a growth in openness towards bidirectional (or multidirectional) formats of museum mediation.

Our approach to conceptualising these conditions of activity and guidance for our prototype was informed by the concept of *conversational agents* from the domain of natural language processing (NLP). The communication here between human and computer is understood as a *dialogue* of two or more agents. Initiative, in this research domain, refers to the concept of "who is in control of the communication" [LiMC15]. We consider that the concept of *initiative* is not limited to dialogues using natural language [Horv99]. It can be extended to other modalities, such as gesture, gaze and movement-based interaction.

Public displays that facilitate the dialogue between human and computer in the domain of human-computer interaction are commonly described as either *autoactive, reactive* or *interactive* systems [Saut04]. In the case of autoactive systems, the computer system performs a monologue during which it holds 100 % of the control over the dialogue (Table 1). Some agency might be given to the human by choosing a *reactive* communication strategy. Examples of a *reactive* system design are interactive videos, an automated call centre or automatic self-opening doors. Still, the degree of agency of the system remains quite high in a reactive system. *Interactive* systems offer the participant of a "communication" the primary control of the dialogue. Nonetheless, the user has just a medium degree of agency, as the system demands a certain behavioural style to follow and decides on the topic of a "dialogue". Examples are office software applications or games.

However, using the HoloLens technology allows us to conceptualise a third mode of activity – the co-active strategy. In the latter, there are times when the system takes or gives the initiative, but, at the same time, the human can interrupt and take the initiative. Both entities, human and computer, form a more natural dialogue, since both can initiate an activity or topic. This mode of activity in NLP is called mixed initiative (Table 1).

From a museum perspective, autoactive and reactive systems would be considered *guided* (in NLP terms system-initiative) and an interactive system *self-guided* (in NLP terms user-initiative). Co-active systems resemble bidirectional mediation in museums (in NLP terms mixed-initiative). Since we want to explore whether and how our interpretation strategy of augmenting a "trained gaze" and the experience of *presence* in our PMR is affected by the mode of activity, we decided to realise all three conditions in our prototype in order to be able to test the effect in an *in situ* user study at a later stage.

| Human-Computer Communication Continuum | | | | |
|---|---|---|---|---|
| *Domain* | *Domain-specific terminology* | | | |
| Museum Mediation | guided | | self-guided | bidirectional |
| Natural Language Processing | system-initiative | | user-initiative | mixed-initiative |
| Human-Computer Interaction | autoactive | reactive | interactive | co-active |

|  | *Examples* | | | |
|---|---|---|---|---|
|  | - 3D Movie<br>- self-playing piano | - interactive video<br>- automatic self-opening door | - software application<br>- arcade slot machine | - natural dialogue systems<br>- adaptive menus |
|  | *Degree of Agency* | | | |
| Human | none | low | medium | dynamic |
| Computer | maximum | high | medium | dynamic |

Table 1: Human-computer communication continuum with domain-specific terminology for the three different modes of activity.

## 3 Visitor Study and Formative Study: User-centred Content Development

We decided to create a prototype for the purpose of our study that exemplifies our concept with a single museum exhibit. The object we selected through a formal object scouting process is the figurative sculpture "Viktoria" from the collection of the Deutsches Historisches Museum (DHM). This 3.90-m tall allegorical representation of the goddess of victory, completed in white marble by the German sculptor Fritz Schaper in 1885, opens up especially two interconnected lines of interpretation which are easy to link with the visual and material properties of the object: the iconographic meaning of a number of attributes and the object's biography and its interconnectedness with German history.

We conducted a two-stage visitor study and a formative study including the HoloLens in order to better understand how visitors interact naturally with the sculpture in an unguided modus and what elements of the sculpture spark their interest. For the visitor study, we observed the visitors' behaviour with and around the sculpture. We randomly selected visitors who entered the area where the sculpture is located and observed and manually noted their movement and behaviour without interacting with them.

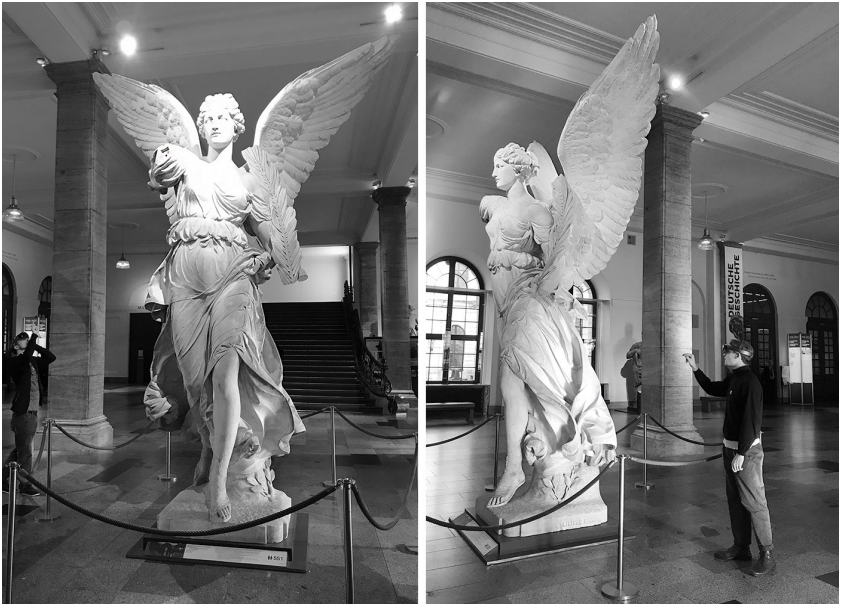

Fig. 2: "Viktoria from the domed room – the Rulers' Hall – of the Berliner Zeughaus" by Fritz Schaper, 1880/1885, DHM, Berlin. Images taken during the creation of the world mesh and three-dimensional model with the HoloLens.

The marble sculpture is placed in the middle of the lower entrance hall of the DHM, approximately 10 m from the ticket counters and 9 m from the staircase that leads to the permanent exhibition. About 30% of visitors passed the sculpture without stopping to look at it; 70% of visitors approached the sculpture and looked at it and about 25% of visitors took a picture of or selfie with the sculpture. After having noted their behaviour, we approached the visitors, including those who had not looked at the sculpture, and asked them to participate in our survey. The observation and survey took place over five days, spanning 3–4 hours on weekdays and weekends in March and April 2019, during which we were able to complete 62 questionnaires. Our main interest was to learn more about the visitors' information needs. We wanted to find out with our first question whether the visitors were able to identify the allegorical figure as the goddess of victory. Out of the 62 respondents, 67% were not able to speak about either the meaning or name of the sculpture, 19.4% of the respondents were able to correctly state the name or the allegorical meaning ("Nike", the Greek equivalent to "Viktoria", or "goddess of victory" were also accepted as a correct answer), while another 12.9% were able to quote the name of the figure because they had just read it on the label.

We then asked the respondents which feature or aspect of the sculpture made them particularly curious. Recurring answers mentioned wings (25%), size (17%), missing arm (9%), overall appearance (6%), robe (5%) and place of installation (5%). When asked what they would like to know more about, the most commonly mentioned aspects were the history or biography of the object (25%), for example, in relation to art history and the history of Berlin, the symbolic meaning of the sculpture or certain attributes (20%), the production process of the artwork (17%), its original place of installation (15%) and details about the visible damage to the sculpture (12%), such as the missing right arm. Interestingly, further information on the artist only made up 6% of the answers, despite authorship and artist attribution being one of the most common pieces of information given on labels and interpretative media in museums.

Having gathered these insights, we then conducted a formative gaze-fixation study with the HoloLens in order to confirm that the self-reported visual aspects that spiked the visitors' curiosity match their actual viewing behaviour as recorded by the HoloLens. Convenience sampled visitors were instructed in putting on the HoloLens and confronted with two tasks: 50% of the participants were told that we wanted to test the wearing comfort of the HoloLens and that they could just walk around the sculpture and look at it for as long as they wanted to. The other half of the participants were told that they had to answer questions about the sculpture at the end of the test.

We were able to confirm that there is a general overlap between the visual features and aspects that visitors had verbally reported as "sparking their curiosity" and the gaze-fixation on the sculpture that we recorded with the HoloLens. In both studies, some important attributes or properties, such as the palm branch and the missing arm, received less attention than we had expected. Furthermore, the gaze-fixation study showed that smaller material aspects, especially those that would require a thorough observation from the side or back of the sculpture, would likely go unnoticed. This included the inscription on the bottom right side of the sculpture's base denoting the artist name and year of creation. The gaze-fixation study also showed that people observed the sculpture more thoroughly when under the impression that they had to answer questions about the sculpture after exploring it (an average of double the time).

Based on the insights from both the self-reported interest and the gaze-fixations recorded with the HoloLens, we defined a total of seven "Regions of Interest" (ROIs) and used them as reference points for the content development process. In this selection, we also included those ROIs that had been missed by visitors during the gaze-fixation study, especially when these ROIs could be linked to the interpretation about recurring self-reported areas of interest derived from our visitor study. We added an introduction and a conclusion to these seven ROIs, resulting in a total of nine content units that can be accessed in the prototype of our PMR experience.

## 4　Content and Design

We generally aimed at creating a PMR that would encourage visitors to thoroughly observe the sculpture, its material presence and attributes. Following our object-centred concept of augmenting a "trained gaze", we made sure that the content linked to our seven ROIs does not distract from the visible elements of the sculpture but reinforces thorough observation. We chose a content type for each of the ROIs that would support the understanding of the associated meaning and its particularity for the sculpture, while also making use of the specific design options of a PMR created with the HoloLens. This led to the decision to use spoken texts to explain the meaning of attributes and material properties of the sculpture. Interpretation in our PMR also includes still images, animations, three-dimensional objects, visual effects and text, each positioned in relation to the sculpture. In the following, we describe the content of the ROIs and how we decided on the selection technique for those. Finally, we detail the different navigation paths which interconnect the ROIs and how they relate to the three different modes of activity.

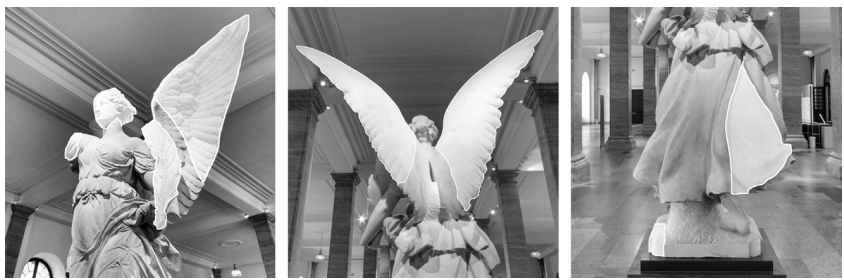

Fig. 3: Regions of interest highlighted in white: missing arm (ROI 1), palm branch (ROI 2), front of the wings (ROI 3), head (ROI 4), back of the wings (ROI 5), inscription on the base (ROI 6) and garment (ROI 7).

## 4.1   Linking Content to Regions of Interest

We are able to augment and, thus, extend the sculpture with the HoloLens by creating a virtual reconstruction of the missing arm (ROI 1) that was originally stretched upwards and held a laurel wreath. The laurel wreath is one of the main attributes of "Viktoria" and supports the identification of the sculpture as the goddess of victory – it already symbolised fame and victory in ancient Rome – which is explained in an audio text. The arm with the laurel wreath was damaged and lost during the Second World War. It was sculpted from an individual piece of marble and attached to the body of the sculpture at the armpit, where it was held in place by a hexagonal mould that is still visible today. The second attribute of "Viktoria" is the palm branch (ROI 2) that she holds in her left arm. When this ROI is selected, a text displays the palm branch's allegorical meaning of victory, triumph and peace, alongside a selection of images from various art historical epochs, allowing the visitor to visually compare the depiction of palm branches. Together with the wings (front of the wings: ROI 3), which trigger an audio text, these two attributes allow the visitor to identify the sculpture as a representation of the goddess of victory. The head (ROI 4) triggers a visual augmentation of the size and weight of the "Viktoria" directly on the object while additional information about the production process is delivered as audio alongside still images that show the assembly in the museum. In contrast to the finely worked marble on the front, the back of the wings (ROI 5) show distinct traces of the chisel that was used by the artist. An audio text related to this ROI interprets the visible difference in the surface texture as a hint that the sculpture was created with a distinct front side and that the back of the wings was not visible at the original location of the sculpture.

Regarding the self-reported interest of visitors to understand the historical relevance of the object in relation to art history better as well as its significance and symbolic function for Berlin, we decided to develop a spatial data visualisation as part of our PMR. A visitor with a "trained gaze" would be able to access a "mental archive of images and style" [HaKl06: 12], which our non-expert visitor would not be able to do. In order to enhance the untrained viewer with a "trained gaze", we had to find a way to place the "Viktoria" in context with other artworks. The visitors should be supported in understanding how the emulation of works of art led to the incorporation of visual characteristics and symbolic meaning known from Greek and Roman antiquity in sculptures created during Prussian Classicism (19th century). To achieve this, a timeline visualisation is rolled out behind Schaper's "Viktoria" in our PMR experience, creating a shared space that the visitors can explore: the physical museum object occupies the space together with a selection of other "Viktorias" and "Nikes", dating from Prussian Classicism as well as Roman and Greek antiquity ("Nike" being the Greek equivalent and predecessor of the Roman "Viktoria"). The two-dimensional images of the related sculptures are placed in three distinct rows on the timeline: one row with artworks created during Prussian Classicism, the second with sculptures dating back to Roman antiquity and a third row with Greek sculptures. The Roman and Greek sculptures are all placed on the timeline according to the year in which they were found or excavated during the 18th and 19th century. Consequently, all sculptures can be observed in a shared historic framework that supports the interpretation that Schaper and his contemporaries took inspiration from and emulated these Greek and Roman sculptures. The timeline is linked to the inscription on the lower right side of the statue's base (ROI 6), which shows 1885 as the year of completion and the artist's name "Schaper", thus, the first mark of the virtual timeline is created by the year 1885 as a visible aspect of the physical object.

Lastly, a fold of the garment at the back of the sculpture was included as ROI 7 with an associated audio text that encourages the visitor to examine the skilfully sculpted folds in the dress closely and to circle around the sculpture in order to follow the figure's movement.

The final view (conclusion) in the PMR experience augments Schaper's "Viktoria" in its original location. Before its partial destruction during the Second World War, the "Zeughaus", the building that nowadays houses the DHM, was used as a military museum. "Viktoria" was placed here at the centre of attention in the "Ruhmeshalle" (Hall of Fame), where she stood on a pedestal in a wall niche at the end of the domed hall. Based on historical photographs, we created a virtual 2.5-dimensional reconstruction of the "Ruhmeshalle". By lowering the virtual floor in the PMR below the actual floor level, it seems like "Viktoria" is again placed on a pedestal in the wall niche. This final view is also accompanied by a conclusion delivered as an audio text. Apart from the introduction and the final view that have fixed positions at the beginning of the experience and after completing all other content units, the content had to be created in such a way that it could be received in a guided, self-guided or bidirectional mode. We had to make sure that all three conditions of initiative were created with exactly the same content in order to be able to accurately analyse the influence of the mode of activity (as described in chapter 2.3) on the learning results and user experience. The content has to lend itself to being received in no particular order in the self-guided and bidirectional mode but also support a smooth and interesting narrative in a guided mode.

### 4.2   Designing the Selection Technique for the ROIs

Apart from the order of the content units, the modes of activity are distinct in their selection technique, which is used to trigger the content. A selection technique typically consists of navigating a sort of pointer (in our case, the sliding puck) to the target, followed by a confirmation that the area currently pointed at should be selected. Many HoloLens applications use the "air tap" gesture [MS20c] as the confirming action, which requires the user to learn this interaction technique [TaAL18]. We decided to not employ "air tap" as an interaction technique in our application to keep the learning phase low and emphasise our gaze-based interaction concept. We initially considered developing a dwell-free selection technique, similar to Sarcar et al. [SaPC13], Porta et al. [PoTu08] or Wobbrock et al. [WRSD07]. However, we found that the currently known and tested dwell-free selection techniques require a similar learning phase to the air tap. Therefore, we decided to fall back on a dwell time-based selection, which has been shown to perform quite well [YGYY17]. Dwell time-based selection techniques typically bear usability problems, such as the Midas touch problem [Jaco95], which means that users are hesitant to look at certain parts of an object because they are afraid of accidentally selecting things. One solution to avoid this is to explicitly design a virtual button attached to the physical object that could be used to confirm selection. However, this does not correspond with our concept of assisting the untrained viewer with a digitally augmented "trained gaze". The search for such buttons would dominate the visual exploration of the sculpture, which, instead, should be focused on the material presence of the object in its entirety. Furthermore, attaching fixed virtual buttons to the sculpture for each content sequence would create a harsh aesthetical interference that we tried to avoid. Instead, the content is selected by first entering a defined collider area around an ROI with the gaze (sliding puck) and a subsequent dwell time-based confirmation (Fig. 5). This means that the visitor's gaze has to remain inside the collider area of the ROI in order to confirm the selection (Fig. 4 right), thus, encouraging thorough observation.

Confirmation constitutes only one part of the two-part selection technique and is similar in all three modes of activity. The second part, navigation, which precedes confirmation, is used to differentiate our modes of activity and guidance, which we describe next. For both parts of the selection technique we established a design solution that would create minimal obstruction and keep the sculpture visible at all times. This solution builds on the decision to use an ephemeral style of particles as visual cues and interaction feedback.

### 4.3   Navigation in three different modes of activity

Navigation can be either user- or system-initiated. In the self-guided, user-initiative mode, visitors have to actively explore the object with the sliding puck attached to their gaze in order to find ROIs with linked interpretation. The sliding puck reacts with the surface of the sculpture and starts to emit particles only when their gaze enters the collider area of an ROI (Fig. 4 left). After two seconds of hovering over the ROI, the associated collider mesh highlights the entirety of the ROI with particles. A dwell timer (2-s long) then forms around the sliding puck and indicates the approaching act of confirmation (Fig. 4 right). When the dwell timer is completed, a selection sound is played and the content associated to the ROI is delivered (see Fig. 5 left). In this self-guided mode, content can be selected in any order, since the user actively explores the statue's features while the system only cues possible links to further information by emitting particles that indicate that the gaze has entered one of the seven ROIs. The visitor can also decide to select content units several times.

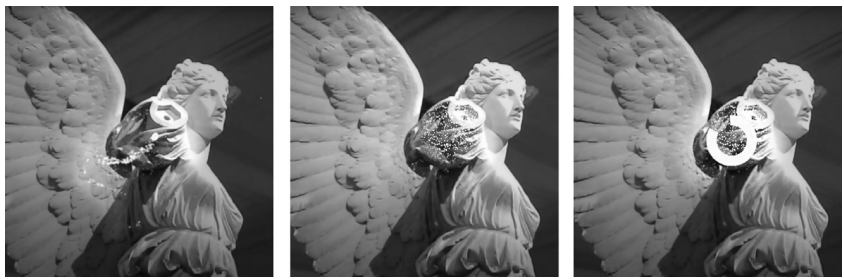

Fig. 4: In the self-guided mode, particles emit from the user-controlled gaze as soon as it enters the collider area of an ROI (left). In the guided mode, the system displays particles over the entirety of the ROI, which has to be confirmed by the user (middle). A dwell timer forms around the sliding puck and has to be completed before the content that is linked to this ROI is delivered (right). As soon as the content is triggered, all particles disappear. Particles are edited and enhanced for visibility in these images.

The agency of the user is reduced in the guided, system-initiative mode of activity, and the system decides what should be selected next. The system initiates the selection process by highlighting an ROI with particles (Fig. 4 middle). If the user follows this visual cue and enters the highlighted ROI with their gaze, the dwell timer becomes visible (Fig. 4 right), and when it is completed, a selection sound is played and the content is delivered (see Fig. 5 right). In this mode, the users can only confirm they have followed the system's cueing, are looking at the ROI and are ready to receive the associated content. The narrative structure of this guided mode is sequential, with a given order and duration. Once played, the content unit cannot be selected again, and the system visually asks the user to confirm the selection of the next ROI by highlighting it with particles.

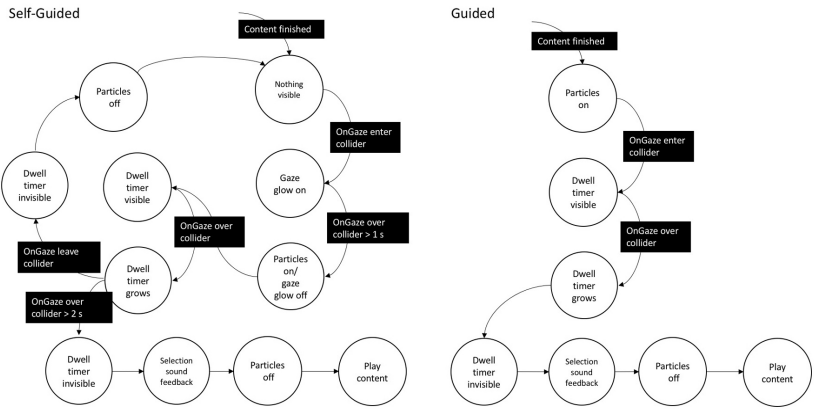

Fig. 5: State diagram showing the initiative design in the self-guided (left) and guided (right) mode.

The bidirectional, mixed-initiative mode uses a mix of both cueing and navigation mechanisms laid out above, depending on who is taking the initiative in the "dialogue"; either the system has the initiative to cue an ROI (system-initiative) or the initiative is given to the user to explore the statue actively (user-initiative). The content structure in this mixed-initiative mode is dynamic and follows a semi-scripted structure. After the introductory audio text, which is automatically played in all three modes, the system firstly switches to the self-guided mode, and the visitor can actively explore the statue and freely select an ROI. In order to be able to initiate a reasoned and purposeful initiative shift, four of the seven content units that we identified as crucial for the understanding of the allegorical and historical meaning of the statue were declared core "learning goals". This included the laurel wreath (ROI 1) and palm branch (ROI 2) – the two main attributes of the goddess of victory – and the head (ROI 4), that is linked to information about the size, production process and assembly of the sculpture in the museum, and, finally, the inscription on the base (ROI 6), that triggers the spatial timeline visualisation (see above). Human and computer have equal amounts of "talking time" throughout the dialogue in the mixed-initiative mode, as initiative switches after each content selection [AlGH99]. The "intention" of the system is to only "talk" (cue ROIs) about the four content sequences that we had declared as important, one at a time. In this bidirectional condition, the system takes the initiative to support the visitor in identifying key concepts, which should subsequently allow the visitor to engage in a more self-determined process of exploration and meaning-making. As soon as all four important ROIs have been received, either by the initiative of the visitor or that of the system, the conclusion sequence is played.

## 5 Discussion and Next Steps

In this paper we hace put a focus on illustrating the concept of our study prototype, which we have based on related research in HCI and museum practice. In order to explore the research questions laid out at the beginning of this paper, we will as a next step conduct an in-situ user study with our prototype. We will explore a variety of research interests or topics with the user study, which are briefly summarised in the following.

First of all, we explore whether the presence of the physical museum object is maintained even when it is augmented with additional content. For this purpose we designed a questionnaire that contains heuristics that aim to help us understand the perception of object presence. Additionally, we want to investigate whether and to what extent the users experienced a sense of "being present" in the PMR application. The concept of presence in the second sense has been a focus in research on virtual reality for almost 40 years. The emergence of presence is likewise commonly studied with the help of questionnaires. However, Usoh et al. [UCAS00], for example, were able to point out the challenge of "measuring" presence through questionnaires. They conducted an experiment that showed that these questionnaires were only useful when all subjects experience the same type of environment, whereas they are not suitable for the comparison of experiences across environments. Although our subjects will all experience the same type of environment (PMR), they will explore the application in three different modes of activity, which may also have an impact on the comparability of the results we derive from our questionnaire. Generally, it will be interesting to see if we can observe a difference in the perceived sense of presence in our PMR, depending on the type of activity, since awareness of possible action patterns has been identified as contributing to a sense of presence [ScFR01]. In addition to the importance of patterns of action, Schubert et al. [ScFR01] point out that attention allocation is also involved in the emergence of presence. While we may not be able to determine the level of presence for our experience through a questionnaire alone, tracking the users' gaze in six degrees of freedom may provide us with valuable insights into whether we were able to guide the visitor to thoroughly explore the museum object in its material presence.

One key aspect that is thought to influence the creation of a pure mixed reality experience is the unbroken holospace (see section 2.1). Nonetheless, the final view, a large-scale 2.5-dimensional reconstruction of the "Ruhmeshalle" that augments the museum object in its original location, consciously ignores this principle. We hope to be able to observe user reactions that allow us to derive indications as to whether or not the fractured holospace in fact reduces the PMR experience.

As we have illustrated in section 2.3, different formats of museum mediation can be understood as prioritising different modes of activity and guidance. In our user study we will explore whether the three corresponding conditions (user-, system- and mixed-initiative) have a noticeable effect on the perceived sense of agency, overall user experience and learning result.

Building on the assumption that museum visitors are driven and influenced by a variety of identity-related motivations [Falk10] [Bond12], our user study also includes a simplified version of Falk's [Falk10] approach towards the allocation of visitor motivation. By including this aspect, we want to explore whether we can observe differences in the study results depending on the visitors' motivation type.

Although we have made a conscious decision to use exactly the same content in all three modes of activity to ensure comparability of results, we are aware that this limitation might not be practical outside a study context. In common museum practice, different modes of activity and guidance would also require specialised and ideally visitor-dependent content. This is related to the notion that the production of meaning and interpretation, especially in contemporary approaches to art history, is understood as emanating from a personal perspective [HaKl06: 20]. Such an understanding also infers that the act of viewing and interpretation – the "trained gaze" – is not only influenced by scholarly knowledge but also by political and social aspects as well as the interest of the beholder. This understanding of interpretation as dynamic, flexible and viewer-dependent encourages us to consider different lines of interpretation and thus offer more diverse content as part of the next iteration. In this way, we will be able to truly exploit the potentials of digitally supported museum interpretation that not only caters to different needs or preferences for guidance and modes of activity but is also able to account for a variety of interests, learning levels, narratives and media.